\documentclass[a4paper,12pt]{article}
\usepackage[T1]{fontenc}
\usepackage[utf8]{inputenc}
\usepackage{lmodern}
\usepackage{fullpage}
\usepackage{setspace}
\usepackage[margin=1in]{geometry}
\usepackage[british,english]{babel}
\title{Profile Analyst: Advanced Job Candidate Matching via Automatic Skills Linking\footnote{
Copyright \copyright\ 2016 Martin A. COLEMAN. This document may be freely shared, copied, transferred and/or re-distributed, in part or in whole, for any purpose and by any means, provided that credit is given. Permanent ID of this document: 3e18e3f5a876984d20f3985ac3df41ad}}
\author{Martin A.~COLEMAN\\Independent Scholar\\Email: martin@mchomenet.com}
\date{9th February, 2016}
\begin{document}
\maketitle{}

\begin{abstract}
The recruitment process is a slow and inefficient one at best, and a potentially ineffective one at worst. Matching candidates to jobs is one thing, but matching candidates with jobs alongside appropriate expectations and taking into account the right aptitude and suitability for any given role is another thing and in this paper we look at a whole new way of matching jobs with potential candidates as well as matching against their overall suitability for a given role.
\end{abstract}

\onehalfspacing
\section*{Introduction}
There is a problem with recruitment in the world which has been recognised all over, and that is simply the fact that job recruitment is unpredictable, ripe with errors, a huge time-sink and costs a lot more money than simply posting a job add on a job board.

From better suited personality types to specific personality disorders, better matching skill sets to company culture fits, finding job candidates for a role can be a real lottery, with companies losing out most of the time. This can be due to an ineffective screening process with recruitment agencies, as well as a fault with the recruitment process conceptually.

Culture fit is the other problem that is encountered. Certain personality types are better suited for certain environments and others not. It is disastrous to bring someone on-board in a company who does not come from the same mindset as others such as their approach to tasks, how they think about the job itself, quiet vs outgoing and the likely perspective that they will bring to the table.

However, there is a proposed solution to address these problems which is not just a cover-up band-aide solution like so many others, but one which may very well turn the entire recruitment process, and industry, upside down. We delve into what this solution is in general and how it would help companies find better suited candidates in much less time and in a much more efficient manner.

\section*{The Problem}
Some candidates may look good on paper but have no personality to bring their skills out. Plus having to continually schedule interviews wastes a lot of company time, all in the hope that they will find a potential employee from the possibly hundreds of applications they have had to distil to just a handful.

Admittedly, this is why recruitment companies are used, to try to deal with weeding out undesirable personalities and to save the staff time on interviews. This usually comes with a hefty price tag, but recruitment is big business \footnote{http://people2people.com.au/blog/recruitment-industry-becoming/}, so we can not argue that it works on a certain scale.

However, a candidate may be great at selling themselves, so they have passed the preliminary personality test and used up the recruitment agency's time as expected. So they are presented to the client company for consideration. This will result in a second interview with the future potential employer as well and may very well pass. Once they make it in and are selected, the candidate will let their guard down, stop acting and their real self will show. This may well be similar to their "interview persona" and will be okay. Other times, a personality clash will occur, they could be trouble makers, narcissists, sociopaths or other undesirables. They could have even lied about all their skills and bluffed the entire interview process, though this can be found out by doing preliminary technical tests in appropriate industries.

So what has happened here is, over a period of a few weeks, a new candidate has come on board, the recruitment company has taken their commission, the candidate turned out to be a dud and the process begins again. This takes its toll on the client company and the recruiter, more-so on the client company for wasting time, money and company resources on training them into the company culture, while the recruitment company takes a small dint in their reputation. Neither outcomes are favourable.

The problem is extended online as job candidates search for roles on job boards, and due the nature of the recruitment industry and low barrier of entry for new providers, one role may be listed up to 20 or more times, giving job seekers the impression of there being more jobs available than there realistically is. This inflates job numbers and gives job hunters false hope. So many job seekers try to exclude recruitment listings from job boards to find the roles that are listed by companies directly, but this is impossible on many popular job boards \footnote{http://forums.whirlpool.net.au/archive/1740104}.

One other problem from a candidate point of view is the application reviewing process. A candidate may apply for a job, might get an automatic responder confirming receipt of their details and may or may not hear back within a few days if they are shortlisted, or may need to wait up to several weeks or months to hear that they did not get through. Candidates then have to play a numbers game of sending out heaps of applications and hoping for the best while not expecting to hear of any news at all, and if they receive news quickly, it's like winning a prize, because communication is not very open at certain agencies.

Granted, that is all a problem with the logistics, but candidates want an honest assessment of the markets such as; [1] how many jobs are really out there, [2] where they stand in the application process, [3] are their skills in demand and [4] how can they be found. An automated control with alert and status functions at all steps along the way would be a \"killer app\" for recruitment, human resources and hiring.

\section*{The Profile Analyst}
An algorithm to automatically link job listings with required skill sets to candidate specifications was explored in theory and in prototype form. This allowed a job listing to be created, with specific skills that were appropriate and the candidate matching system matched the number of skills that matched against a candidate that the client company was looking for. This was then presented as a list to the client after which they were able shortlist potential candidates. It used a percentage system that informed the client how likely they were to be the ideal candidate for any particular job listing.

The data that was collected from a new candidate user included first name, last name, email, date of birth, location, minimum income of pay range required, maximum income of pay range required, employment type (full time, part time, casual or contract), personality type and a profile photo.

The data that was collected from an employer was simply the company or business name, HR contact name, phone, email and logo.

Within the job listing itself were linked the to employer internal ID, job title, ideal age, ideal gender, summary of the role, offered salary of position, skills required and the ideal personality type.

The personality type was closely, but not entirely, modelled on the Myers-Briggs Type Indicator. This meant that it had to distinguish between outgoing or more reserved people, those who had more empathy versus those who governed entirely from their mind, whether someone was a team player or more of an autonomous lone wolf, whether someone was likely to reject authority or take commands more easily and whether they needed strict routine in their work or if they might be more flexible in their approach.

A skills list was also created that would be a collective list of many of the most recognised skill sets from the most commonly accepted categories of industries.

When the candidate first signed up, they would answer the preliminary questions, then be presented with a quiz of 25 questions in the form of a personality best-fit aptitude test. This would determine which theoretical type the candidate would be and link it to their profile internally to aide the profile analyst job matching algorithm. They would be include their skills list within their profile.

A controversial point was brought up that such a system would implicitly allow discrimination against age, gender and appearance but this point was generally accepted to a real world problem that occurred regardless, even against everyone's best intentions. However, this could have been partially dealt with by giving a certain weight scale to these criteria so that age or gender would only have a 50 percent decider on a role. This would help with being gender neutral for mixed roles such as retail, hospitality or teaching, but might be useful in reception work or gardening and maintenance roles.

This therefore gives us several tables so far; Candidate, Employer, Job, Skills, Personality Result, Personality Questions, Personality Answers and Shortlist. The shortlist was generated in memory, triggered by each posted job opportunity and then kept track of in 4 stages: [1] Whether the employer had shortlisted the potential employee, [2] Whether the potential employee had shortlisted the potential employer, [3] Whether the employer had attempted to initiate contact with the potential employee and [4] Whether the potential employee had responded to the contact request. Once all 4 conditions were met, the system allowed instant messaging and later, video conferencing between the two parties.

\section*{Making It Work}
The overall design includes 3 aspects: [1] The front end. This could be web-based or a mobile application, [2] The information processor (this is in 2 sub-parts) and [3] The database.

\textbf{1. The Front End}\\
This is quite simple. A User Interface built around the idea of passing commands and requests to the server (in the case of the prototype, via JSON), retrieving the information and presenting it to the user. Both job seekers and businesses would be able to register an account and upload their details.

Job seekers would enter their skills and basic employment history (essentially making the idea of a resume redundant) and upload a profile photo. They would then do the personality and aptitude test and the results are processed by the server.

Businesses would use the app to list a job, along with the skill sets requirement, personality type that would be the ideal fit as well as industry, job role/title and offered pay rate. Following that, the system would generate a list of suitable candidates automatically, starting with the ones most suited taking into account all the other criteria (location, skills, personality, pay wanted, gender) all the way to ones that do have some skills matching but may not be best suited.

HR and hiring managers would then be able to go through the list and choose, either by swiping or by checkbox, which of the candidates they wish to shortlist.

Once they have shortlisted, they may make a contact request with the candidate. Once the candidate accepts the request, they contact each other via instant messenger to set up a video chat for a quick interview. How the rest of the interview process goes with that candidate is then between the HR contact and the candidate.

\textbf{2.1 The API}\\
The API which acts as the go between for the GUI front-end and the database accepts all commands and sends back requested data via JSON. Basic tests using normal POST commands with CSV results were used and later it was converted to JSON for simplicity between platforms.

Command requests included user registration, user profile updating, personality assessment, accept contact request and accept video request. It also allowed business registration, HR contact registration, business profile updating, listing of jobs, fetching a list of viable candidates and keeping track of the shortlisting.

\textbf{2.2 The Profile Analyser}\\
This is where the \"magic\" happens. The profile analyst is triggered whenever a new job is listed. It performs a search on the Candidates and Skills database to match those who have the skills that fit the job role and who have a salary requirement within the offered range and/or if the candidate has that part \"open\" (ie: can be negotiated).

A basic search via SQL would look similar to: SELECT cID, cFName, cLName, skCID, skName FROM Candidate, Skills JOIN Skills GROUP BY Skills.

Even though it would be possible to apply the additional logic to the results via SQL, it is best to exclude as much business logic from the database as possible as the Profile Analyst should never be dependent upon the one database in case later design decisions dictate an entirely different database model (MongoDB, NoSQL, etc). So we only want to do a preliminary search and sort in SQL and then perform the core in-depth analysing and weighting in code.

The results are brought back into an array where we may use quicksort or hashtables to get the results we want and then feed them back to the client via JSON. This is left as an exercise of the reader for how to implement.

\textbf{3. The Database}\\
The database is probably one of the simplest aspects. Using a relational database in the prototype, this holds a carefully designed set of tables to hold the information for the Candidates, Companies, Jobs, Skills, Shortlisting. It needed to be established on an open standard so it would be flexible, transferable (from single server hosting to cloud easily), allow fail-over redundancy and duplication.

\section*{Discussion}
Another feature of such a system would be that there is no searching for candidates. The hypothetical Profile Analyst would generate a list of viable candidates as soon as a job listing is opened and would create a feed of suitably skilled candidates with appropriate personality types automatically. This would happen for both the client companies and the job seekers.

Client companies would be able to browse the results, finding out what their expected pay is, their geographic location, their gender, what they look like, and their basic employment history. Gender and appearance is a sensitive issue, but one that has real world application. One basic line of reasoning is that women are more empathetic, sensitive and friendly to strangers, so they are more suited to reception work and office management whereas men typically from a sociological viewpoint are not as much. In turn, it is men who would be more likely the ones to be seen doing construction work on roads and buildings. This is the unspoken but real world gender stereotyping that occurs and has to be worked with in order to make a practical system.

\section*{Conclusion and Further Development}
As ambitious, efficient and practical as the above results may show, there are still further tweaks that may improve the job matching results dramatically. The above process is greatly improve job matching systems in leaps and bounds but one further improvement would be to include the ability to match the most appropriate personality type.

Myers-Briggs Type Indicator (MBTI) has long been used as one of, but definitely not the only, primary methods of judging the personality type of the test taker and then finding which career and/or industry would be most appropriate for that personality type\footnote{https://www.personalitypage.com/html/careers.html}. One good example is that an introverted person would not be ideal for an outgoing sales role that requires a lot of outward confidence, being a persuasive speaker and constantly engaging in enthusiastic conversation regarding products or services. However, this introverted person would be idea as a systems analyst, if they showed the aptitude for it, or even a doctor or journalist.

So matching personality types that are better known for working in specific roles within companies and within certain cultures without having to go through probation only to find out the candidate is secretly miserable would save an immense amount of time, effort and resources.

Of course, one does not necessarily need to think about something as drastic as completely shaking up the recruitment industry (though an argument might be made that it is necessary), the same formulas could be used in recruitment itself to greatly improve matches within candidate databases. The methods used by iProfile, LinkedIn and Seek, to name a few, are horribly ineffective, inefficient, rely upon everyone being on the same page as to terms and lingo to look out for (which is rarely the case) and are so generic that recommendations land in a users inbox which are in no way relevant than simply being in the same industry (and many industries are incredibly vast!).

\end{document}